\documentclass{article}
\pdfoutput=1
\usepackage[letterpaper, portrait, margin=1.5in]{geometry}
\usepackage[T1]{fontenc}
\usepackage{color}
\usepackage{nicefrac}
\usepackage{multirow}
\usepackage{float}
\usepackage[english]{babel}
\usepackage{amsmath, amssymb, mathtools}
\usepackage{fancyhdr}
\usepackage{scrextend}

\title{\textbf{The Quantum Prediction for Einstein-Podolsky-Rosen (EPR) Experiments}}
\author{Donald A. Graft\\\textit{donald.graft@cantab.net}}
\date{}

\DeclarePairedDelimiter\bra{\langle}{\rvert}
\DeclarePairedDelimiter\ket{\lvert}{\rangle}

\begin{document}
\lfoot{}
\cfoot{}
\rfoot{}
\lhead{}
\chead{}
\rhead{}
\maketitle
\thispagestyle{fancy}

\begin{abstract}
Quantum mechanics allows for multiple predictions for the outcome of an EPR experiment. The correct calculation must be used, guided by the physical conditions of the experiment. The quantum joint prediction for EPR correlation is derived and shown to involve a single sampling. The solution for separated measurement, where there are two private samplings, is then developed using orthodox quantum mechanics with L{\"u}ders' rule for state projection on measurement. The separated solution is shown to duplicate the predictions of the quantum joint solution. However, it is shown that this solution requires superluminal transmission of information and therefore it is physically impossible. Alternative predictions (including no correlation) are developed using both Von Neumann projection and null projection (no projection at all). Conditions for the proper application of state projection rules are considered and L{\"u}ders' rule is identified as the primary culprit in the EPR paradox. It cannot be applied to EPR.
\end{abstract}

\newpage
\rhead{\thepage}

\section{Motivation}
\label{Motivation}

In earlier papers  \cite{Graft00,Graft01} I argued that the orthodox quantum mechanics solution for EPR is incorrect and that an alternative calculation is required. My previous treatment, although correct, was based on a macroscopic, classical model (spinning disks). I argued that it fully embodies the logic of the quantum solution because it delivers probabilities and correlations predicted by quantum mechanics for joint sampling, and can therefore be useful in explicating issues of separation. However, the argument can be dismissed by those not disposed to see the logical equivalence, or those demanding a purely quantum approach. The previous treatment, by starting with the final quantum eigenvalue probabilities, missed the chance to show the important role of quantum state projection. Here I fill this lacuna of my previous treatment by using quantum mechanics from the beginning.

A second important motivation is to demonstrate that there is not one unique quantum mechanics prediction for EPR. The literature almost universally refers to ``\textit{the} quantum mechanics prediction'', as if application of L{\"u}ders' extension of Von Neumann's projection postulate is the only possible calculation. In fact, however, as shown here, there are several possible  calculations, corresponding to the projection rule that applies to the physics of the experiment.

The scheme of the paper is as follows. The quantum joint prediction for EPR correlation is derived and shown to involve a single sampling. The solution for separated measurement, where there are two samplings, is then developed using orthodox quantum mechanics with L{\"u}ders' rule for state projection on measurement. The separated solution is shown to duplicate the predictions of the quantum joint solution. However, it is shown that this solution requires superluminal transmission of information and therefore is physically impossible. Alternative predictions respecting special relativity are developed using both Von Neumann projection and no projection at all. Conditions for the proper application of state projection rules are considered and L{\"u}ders' rule is identified as the primary culprit in the EPR paradox. It cannot be applied to EPR.

\section{The EPR paradox}
\label{The EPR paradox}

Difficulties in the foundations of quantum mechanics first exposed by Einstein, Podolsky, and Rosen~\cite{Einstein00} continue to bedevil our understanding of composite systems with spatiotemporally separated components. The cloud of paradoxical considerations that accompany these difficulties can be termed `the EPR paradox'. The EPR paradox has diverse facets and sequellae but our purposes here allow us to state matters simply as follows. Consider a pair of spin-$\nicefrac{1}{2}$ particles in the (anticorrelated) singlet state. When the particles are measured separately at stations A and B for their spin values along two arbitrarily chosen directions $a$ and $b$, the correlation of the two results is predicted to be $-cos(a-b)$, no matter how far apart spatially the measurements may be performed. Experiments are said to confirm this prediction. However, it is easy to show that no local model can produce such correlations~\cite{Bell00}. If nonlocal effects are introduced, superluminal transmission is required and Lorenz invariance is violated. That is the paradox. Einstein famously called the postulated nonlocal effect ``spooky'', Schr{\"o}dinger called it ``sinister'', and Margenau called it ``monstrous''.

Notwithstanding the apparent horror, the current consensus is that the paradox is resolved by accepting the existence of 'quantum nonlocality', whereby the correlations are explained. Concerns about the conflict with special relativity are set aside by right of the no-signaling theorem. I show that such expedients are superfluous and that the paradox can be dissolved without appealing to nonlocality. I also argue that the concerns about a conflict with special relativity cannot be set aside.

\section{The quantum joint prediction for an EPR experiment}
\label{The quantum joint prediction for an EPR experiment}

Here I develop the quantum joint prediction for an EPR experiment and show that it involves a single sampling. This derivation is typical of and consistent with the derivations invariably encountered in support of quantum nonlocality.

Consider a source emitting pairs of spin-$\nicefrac{1}{2}$ particles in the (anticorrelated) singlet state. The $4x4$ density matrix for this state is:

\begin{equation}
\begin{aligned}
\rho_{singlet}^{4x4} = \frac{1}{2}\begin{bmatrix} 0 & 0 & 0 & 0 \\ 0 & 1 & -1 & 0 \\ 0 & -1 & 1 & 0 \\ 0 & 0 & 0 & 0 \end{bmatrix}
\end{aligned}
\end{equation}

Note that here and throughout the paper, density matrices are given in the Z representation, and the particles are propagating along the Y axis. Measurement angles are chosen relative to the Z axis in the X-Z plane. 

The particles separate and are measured at the two stations A and B, configured respectively with the measurement angles $a$ and $b$. The measurement operators measure the outcomes along the measurement axis and are given by:

\begin{equation}
\begin{aligned}
M_A^{2x2} = \begin{bmatrix} cos(a) & sin(a) \\ sin(a) & -cos(a) \end{bmatrix}
\end{aligned}
\end{equation}

\begin{equation}
\begin{aligned}
M_B^{2x2} = \begin{bmatrix} cos(b) & sin(b) \\ sin(b) & -cos(b) \end{bmatrix}
\end{aligned}
\end{equation}

Now, to get the expectation of the product of the outcomes, orthodox quantum mechanics creates a tensor product of the two measurement operators, then applies the resulting $4x4$ operator to the input singlet state, and finally takes the trace of the result. This is represented as follows and a simple calculation shows the result to be $-cos(a-b)$:

\begin{equation}
\langle AB \rangle = Tr([M_A^{2x2}{\otimes}M_B^{2x2}][\rho_{singlet}^{4x4}]) = -cos(a-b)
\end{equation}

\smallskip
Note that the dependence of the trace on only the difference of the two measurement angles implies rotational invariance.

Dephasing of $\rho_{singlet}^{4x4}$ (decoherence) may occur, and if it occurs the EPR correlations may be washed out. I neglect this effect here and assume that coherence is maintained to focus on the operator separation.

The expectation values for the individual results at the two sides are also given by orthodox quantum mechanics:

\begin{equation}
\langle A \rangle = Tr([M_A^{2x2}{\otimes}\textbf{1}^{2x2}][\rho_{singlet}^{4x4}]) = 0
\end{equation}

\begin{equation}
\langle B \rangle = Tr([\textbf{1}^{2x2}{\otimes}M_B^{2x2}][\rho_{singlet}^{4x4}]) = 0
\end{equation}

Here the $2x2$ measurement operators are expanded to $4x4$ operators for application to the $4x4$ singlet by taking the tensor product of the operator and the $2x2$ identity matrix, per orthodox practice.

Because both $\langle A \rangle$ and $\langle B \rangle$ are equal to 0, the expectation $\langle AB \rangle$ gives the correlation between the two measurement outcomes, each of which can yield the value -1 or 1 (the eigenvalues of the measurement operators).

From these expectation values, it is easy to calculate the probabilities of the eigenvalues of the joint outcomes. For example, the probability of both measurements yielding the value 1 is given by $P_{A = 1,B = 1}$. The four probabilities are found to be:

\begin{equation}
P_{A = -1,B = -1} = P_{A = 1,B = 1} = (1 + \langle AB \rangle) / 4
\end{equation}
\begin{equation}
P_{A = -1,B = 1} = P_{A = 1,B = -1} = (1 - \langle AB \rangle) / 4
\end{equation}

My previous analysis \cite{Graft00} started here with a classical system generating these same probabilities.

To generate actual outcomes that can be compared to those of an EPR experiment, sample the probability distribution $P_{i,j}$:

\begin{equation}
Sample(P_{i,j}) \implies O_A, O_B \in 1, -1
\end{equation}

For a large number of source emissions $N$ in an EPR experiment, collect the series of outcomes from each side and correlate them, as follows:

\begin{equation}
{\langle AB \rangle}_{exp} = \frac{1}{N} \sum_{n=1}^{N} {O_A}_n{O_B}_n \approx -cos(a-b)
\end{equation}

The resulting experimental correlation yields the predicted value $-cos(a-b)$. This result is demonstrated by a simulation model that reproduces all the steps given above (with $N = 10000$). The results of the simulation are shown in Figure 1. The measurement angle $b$ at side B is scanned over {$0-2\pi$} while holding $a$ constant. Two values of $a$ are depicted to demonstrate the rotational invariance. The simulation source code is available on-line \cite{Graft02}.

Two important things to note about the quantum joint prediction are that it involves only one sampling, equation (9), and that this sampling relies upon a tensor product of the two measurement operators so that the sampling has access to both of the selected measurement angles. In an EPR experiment, however, there are two samplings, each having access to only the local measurement angle. The joint prediction therefore cannot be used for EPR and we seek a separated solution that properly represents the two private samplings.
\newpage
\centerline {Figure 1. Correlation resulting from quantum joint}
\centerline {measurement in a simulated EPR experiment}
\begin{center}\includegraphics[scale=0.3]{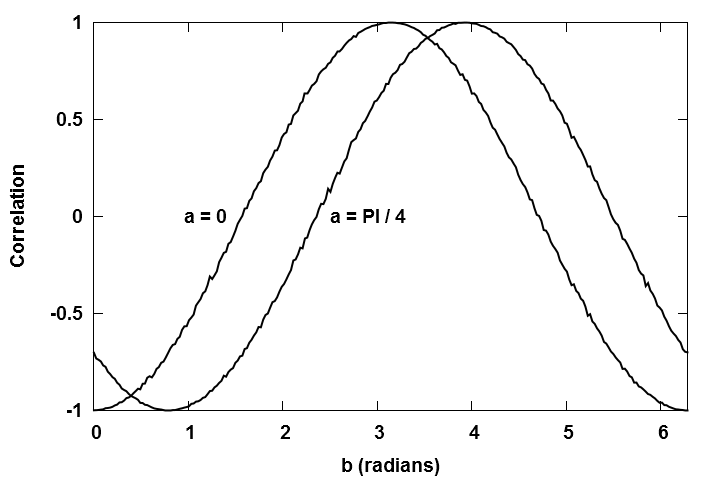}\end{center}

\section{Separating the quantum joint prediction using L{\"u}ders'\\rule}
\label{Separating the quantum joint prediction using L{\"u}ders' projection}

To separate the quantum joint solution we need to account for two samplings each restricted to knowledge of the local measurement angle. We can do this by considering two sequential measurements, specifically, a measurement at side A using the measurement angle $a$, followed by a measurement at side B using $b$. In the degenerate case of the measurements occurring at exactly the same time, the measurement order is chosen arbitrarily. Refer to Section \ref{The separated solution using L{\"u}ders' projection requires superluminal transmission of information; it cannot be used for EPR}
for further discussion of the relativistic aspects.

I now give the orthodox quantum solution for successive measurements in an EPR experiment. Recall the input singlet state and measurement operators previously described in equations (1), (2), and (3). For the first measurement A, the expectation value $\langle A \rangle$ is given by:

\begin{equation}
\langle A \rangle = Tr([M_A^{2x2}{\otimes}\textbf{1}^{2x2}][\rho_{singlet}^{4x4}]) = 0
\end{equation}

Because the expectation value is 0, the probabilities for our measurement results are given by: 

\begin{equation}
P_{A = -1} = P_{A = 1} = 0.5
\end{equation}

Sampling the distribution $P_{i}$ now yields an outcome for the measurement at A:

\begin{equation}
Sample(P_{i}) \implies O_A \in 1, -1
\end{equation}

Now apply the orthodox state projection postulate in the form of L{\"u}ders' rule \cite{Luders}. The input singlet state is rotationally invariant, so it can be expressed in the $a$ measurement basis as follows:

\begin{equation}
\psi_{singlet} = \frac{1}{\sqrt{2}}(\ket{-1_a,1_a} - \ket{1_a,-1_a})
\end{equation}

The $a$ subscripts denote the $a$ measurement basis. If the measurement at A produces a value of $-1_a$, then L{\"u}ders' rule gives the renormalized projected state as:

\begin{equation}
\psi_{L} = \ket{-1_a,1_a}
\end{equation}

This is separable so the projected B state is $\ket{1_a}$, and the corresponding $2x2$ density matrix is $\ket{1_a}\bra{1_a}$. Therefore the projected density matrix in the Z basis for the case of $O_A = -1$ is given by:

\begin{equation}
O_A = -1:~~\rho_B^{2x2} = \ket{1_a}\bra{1_a} = \begin{bmatrix} cos^2(a/2) & sin(a/2)cos(a/2) \\ sin(a/2)cos(a/2) & sin^2(a/2) \end{bmatrix}
\end{equation}

This projection is derived by straightforward application of the orthodox L{\"u}ders rule, also sometimes referred to as the generalized Born rule, especially in the field of quantum information. There is much more to say about state projection rules, but for now we follow the orthodox approach.

Expand $\rho_B^{2x2}$ to $\rho_B^{4x4}$. This is done so that the input state for the second measurement can be assigned as the original $4x4$ singlet state when representing  the case of no projection. The expanded density matrix is renormalized to preserve a unity trace.

\begin{equation}
O_A = -1:~~\rho_B^{4x4} = \frac{1}{2}(\textbf{1}^{2x2}{\otimes}\rho_B^{2x2})
\end{equation}

Expand the tensor product:

\small
\begin{equation}
\begin{multlined}
O_A = -1:~~\rho_B^{4x4} =\\
\\
\frac{1}{2}\begin{bmatrix} cos^2(a/2) & sin(a/2)cos(a/2) & 0 & 0\\ sin(a/2)cos(a/2) & sin^2(a/2) & 0 & 0 \\ 0 & 0 & cos^2(a/2) & sin(a/2)cos(a/2) \\ 0 & 0 & sin(a/2)cos(a/2) & sin^2(a/2)
\end{bmatrix}
\end{multlined}
\end{equation}
\normalsize

Similarly, the expression for the projected state when the measurement at side A yields a value of $1_a$ is given by:

\small
\begin{equation}
\begin{multlined}
O_A = 1:~~\rho_B^{4x4} =\\
\\
\frac{1}{2}\begin{bmatrix} sin^2(a/2) & -sin(a/2)cos(a/2) & 0 & 0\\ -sin(a/2)cos(a/2) & cos^2(a/2) & 0 & 0 \\ 0 & 0 & sin^2(a/2) & -sin(a/2)cos(a/2) \\ 0 & 0 & -sin(a/2)cos(a/2) & cos^2(a/2)
\end{bmatrix}
\end{multlined}
\end{equation}
\normalsize

The expectation value of the result at side B can now be given in terms of the projected state:

\begin{equation}
\langle B \rangle = Tr([\textbf{1}^{2x2}{\otimes}M_B^{2x2}][\rho_B^{4x4}])
\end{equation}

The probabilities for the B measurement result are given by: 

\begin{equation}
P_{B = 1} = \frac{1 + \langle B \rangle}{2}
\end{equation}
\begin{equation}
P_{B = -1} = \frac{1 - \langle B \rangle}{2}
\end{equation}

Sampling the distribution $P_j$ now yields an outcome for the measurement at B:

\begin{equation}
Sample(P_{j}) \implies O_B \in 1, -1
\end{equation}

For a large number of source emissions $N$ in an EPR experiment, collect the series of outcomes from each side and correlate them, as follows:

\begin{equation}
{\langle AB \rangle}_{exp} = \frac{1}{N} \sum_{n=1}^{N} {O_A}_n{O_B}_n \approx -cos(a-b)
\end{equation}

Again, the resulting experimental correlation yields the predicted value $-cos(a-b)$. This result is demonstrated by a simulation model that reproduces all the steps given above (with $N = 10000$). The results of the simulation are equivalent to those shown in Figure 1, with differences due only to stochasticity. The simulation source code is available on-line \cite{Graft03}. The form of projection to be used is selectable, that is, between L{\"u}ders projection, Von Neumann projection, and no projection (see below).

The separated solution successfully combines two private samplings to deliver the EPR correlations. The solution appeals to the orthodox  L{\"u}ders rule of state projection to generate the projected state. However, all is not well.

\section{The separated solution using L{\"u}ders' projection requires superluminal transmission of information; it cannot be used for EPR}
\label{The separated solution using L{\"u}ders' projection requires superluminal transmission of information; it cannot be used for EPR}

Here I consider only the case where the two sides share an inertial frame and make their measurements in that inertial frame. It would be possible to consider relatively moving frames for the two measurement stations, however, nobody claims that quantum entanglement requires relatively moving frames, so allowing for it is an unnecessary complication. It may be interesting to go further and analyze the effects of relativistic motion, but it is not directly germane to the argument here.

In our context, there are always three operational domains relating the measurement times at the two sides (taking the measurement time to be the time of its completion if it is extended in time). In the first domain, measurement A precedes measurement B in the shared frame. In the second domain, measurement B precedes measurement A in the shared frame. In these two domains denote the first measurement by $t_0$ and the second by $t_1$. In the third domain, the measurements are simultaneous, to a certain experimental precision. In this domain, for the purposes of calculation, arbitrarily designate one side's measurement time as $t_0$ and the other's as $t_1$. It is interesting that quantum mechanics has nothing to say about which side projects the other in the case of simultaneous measurements, and I accordingly do not speculate on how the symmetry is broken physically.

In the following material I refer to the measurement times $t_0$ and $t_1$ defined above. It is important to realize that a case of simultaneous measurements (as defined) is not necessarily a case of joint measurement yielding entangled statistics. It is possible to perform marginal measurements simultaneously. This is a commonly overlooked.
\newpage
Orthodox quantum theory employing L{\"u}ders projection, as demonstrated in Section \ref{Separating the quantum joint prediction using L{\"u}ders' projection} for the case of EPR with separation, entails that the measurement at side A produces a projected state at side B. The projected state is one of the two states given in equations (18) and (19). The projected state must be locally present at side B, to be available for the local measurement at side B.

It is obvious from perusal of equations (18) and (19) that the projected state contains information about both the measurement angle $a$ at side A ($a$ appears directly in the density matrices) and information about the outcome (the outcome determines the signs in the entries of the density matrices). Therefore, in terminology more typical for philosophical discussions of quantum foundations, both parameter independence and outcome independence are violated by the separated quantum solution for EPR. Earlier in the history of the EPR paradox it was thought that violation of either parameter \textit{or} outcome independence could suffice to account for the form of the EPR correlations, and much ink was spilled on debating which of the two was operative in EPR. However, it is now clear, not only from the derivation of Section \ref{Separating the quantum joint prediction using L{\"u}ders' projection}, but from other analyses \cite{Graft00,Graft01,Pawlowski}, that \textit{both} parameter and outcome independence must be violated to account for the EPR correlations.

Transmission of information is not problematic as long as the transmission is not required to occur at superluminal speeds, because special relativity would be violated, and that is something that theorists must eschew to maintain a consistent axiom set for physics. However, it is easy to see that the quantum separated solution using L{\"u}ders' rule requires superluminal transmission in the paradigmatic EPR experiment with a large separation between the measurement sides.

Suppose that the measurement sides A and B are separated in space by a very large amount. We can arbitrarily suppose that the separation is one light year (any large separation suffices). Consider without loss of generality the case where side A's measurement time is designated as $t_0$ (we could think of side B measuring at time $t_0$ instead; the general argument is not changed). Suppose now that at time $t_0$ side A randomly chooses a measurement angle and performs the measurement, producing an outcome. According to special relativity, due to the physical separation, the information about the measurement angle and the outcome cannot appear at side B until at least a year has elapsed. However, the measurement at side B can be performed at an arbitrarily short time $t_1$ after the side A measurement at time $t_0$. The timing is determined by the location of the pair emission source between the two sides; we can place it so that $t_1 - t_0$ is arbitrarily small. 

Now, if the EPR correlations are to be obtained in such a scenario, as the current consensus believes, then parameter and outcome information become available in the projected state localized to side B within the arbitrarily small inter-measurement time, which is much shorter than the required time (one year) to physically transmit the information. Therefore, the general separated solution using L{\"u}ders' rule requires superluminal transmission of information, violating special relativity.

One could argue that, although the information is indeed present at side B at time $t_1$, there is no way for side B to extract it (per the no-signaling theorem), but this is irrelevant \cite{no-signaling}, and I emphasize that the information \textit{must have been present} at side B at time $t_1$.

While special relativity prevents us from applying L{\"u}ders projection to EPR, there are other physical scenarios where it may be applied, and so there is no outright prohibition from applying L{\"u}ders projection to all physical scenarios. Several scenarios can be conceived where L{\"u}ders projection may be validly applied (the list is not intended to be exhaustive):
\newpage
\begin{itemize}
\item Separable source states. Information transfer is not needed for separable source states. In EPR, the source state is not separable.

\item Nondegenerate spectra. L{\"u}ders' rule extends Von Neumann's projection rule to  account for degenerate spectra. If there is no degeneracy, then L{\"u}ders' rule collapses to Von Neumann's rule. In EPR, we are dealing with degenerate spectra.

\item Decomposition/analysis of a true joint measurement. If we have a scenario such as the one in Section \ref{The quantum joint prediction for an EPR experiment}, we could apply an alternative analysis in terms of L{\"u}ders' rule, although the usefulness of such an analysis is not clear, as we should prefer to faithfully represent the samplings defined by the physics of the experiment. In EPR, we do not have a true joint measurement.

\item Cases where the physical arrangement does not require superluminal speeds. For example, if our measurement stations are very close to one another, we could consistently assert that subluminal information transmission occurs, in which case we should be able to identify a physical mechanism for the transmission. In the general case of EPR, however, we deal with large separations, such that if transmission occurs, it must be superluminal.
\end{itemize}

Before considering alternative quantum predictions for EPR that satisfy special relativity, I point out that it is possible to empirically test whether L{\"u}ders projection actually occurs in an experimental arrangement \cite{Adler,Hegerfeldt}. Hegerfeldt and Sala Mayato \cite{Hegerfeldt} argue that different forms of projection ``may appear naturally, depending on the realization of a particular measurement apparatus'' and ``Their applicability depends on the circumstances, i.e., the details of the measurement apparatus.'' The fact that actual determination by experiment of the applicable form of projection has never been performed for EPR is surprising and can be viewed as a disquieting state of affairs in quantum foundations research.

\section{Alternative quantum predictions for correlation in an EPR experiment}
\label{Alternative quantum predictions for correlation in an EPR experiment}

\subsection{Von Neumann projection}
\label{Von Neumann projection}

As shown, L{\"u}ders' rule cannot be applied to EPR experiments, leading us to search for alternatives that can yield a correct separated solution. Von Neumann's projection rule produces a mixture upon measurement \cite{VNnote}, while L{\"u}ders' rule produces a pure state. In general the pure state may be a superposition but in EPR the L{\"u}ders-projected Hilbert subspace contains only a single eigenvalue. In our context, Von Neumann projection produces a mixture of the states $\ket{1_a}$ and $\ket{-1_a}$. The separated solution then proceeds as follows.

Recall the input singlet state and measurement operators previously described in equations (1), (2), and (3). For the first measurement A, the expectation value $\langle A \rangle$ is given (as shown before in Section \ref{The quantum joint prediction for an EPR experiment}) by:

\begin{equation}
\langle A \rangle = Tr([M_A^{2x2}{\otimes}\textbf{1}^{2x2}][\rho_{singlet}^{4x4}]) = 0
\end{equation}

Because the expectation value is 0, the probabilities for our measurement results are given by:

\begin{equation}
P_{A = -1} = P_{A = 1} = 0.5
\end{equation}

Sampling the distribution $P_i$ now yields an outcome for the measurement at A:

\begin{equation}
Sample(P_{i}) \implies O_A \in 1, -1
\end{equation}

The $4x4$ density matrix for the projected state at B applying Von Neumann projection is a mixture of the two possible L{\"u}ders' rule eigenvalues given by equations (18) and (19):

\begin{equation}
\rho_B^{4x4} = \frac{1}{4}\begin{bmatrix} 1 & 0 & 0 & 0\\ 0 & 1 & 0 & 0 \\ 0 & 0 & 1 & 0 \\ 0 & 0 & 0 & 1
\end{bmatrix}
\end{equation}

Von Neumann projection in the case of EPR creates a mixture equivalent to a normalized identity matrix.

The expectation value of the result at side B can now be given in terms of the projected state:

\begin{equation}
\langle B \rangle = Tr([\textbf{1}^{2x2}{\otimes}M_B^{2x2}][\rho_B^{4x4}]) = 0
\end{equation}

The probabilities for the B measurement result are given by:

\begin{equation}
P_{B = -1} = P_{B = 1} = 0.5
\end{equation}

Sampling the distribution $P_j$ yields an outcome for the measurement at B:

\begin{equation}
Sample(P_{j}) \implies O_B \in 1, -1
\end{equation}

For a large number of source emissions $N$ in an EPR experiment, collect the series of outcomes from each side and correlate them, as follows:

\begin{equation}
{\langle AB \rangle}_{exp} = \frac{1}{N} \sum_{n=1}^{N} {O_A}_n{O_B}_n \approx 0
\end{equation}

The prediction applying Von Neumann projection is that no correlation will be observed. This calculation is confirmed by a simulation whose results are shown in Figure 2. The source code of the simulation is available on-line \cite{Graft03}. The form of projection to be used is selectable, i.e., between L{\"u}ders projection, Von Neumann projection, and no projection.

At first it seems that we can say that no information (specifically, information about the measurement angle and outcome) is transferred by Von Neumann projection, and that therefore special relativity cannot be violated even for very large separations. However, one can see that at least one bit of information must be transferred to signal projection of the singlet state to the normalized $4x4$ identity matrix, as required. Operationally, side B cannot distinguish a singlet state from a normalized $4x4$ identity matrix by its measurement (satisfying no-signaling), so we might be inclined to dismiss this as inconsequential, and console ourselves with the fact that Von Neumann projection nevertheless gives the correct prediction for EPR (no correlation). Theoretically, however, even one bit of information transferred superluminally forces us to reject Von Neumann projection for EPR. The only correct solution for EPR must exclude all information transfer, that is, it must exclude projection completely.

\bigskip
\centerline {Figure 2. Correlation resulting from quantum separated measurement}
\centerline {using Von Neumann projection in a simulated EPR experiment}
\begin{center}\includegraphics[scale=0.3]{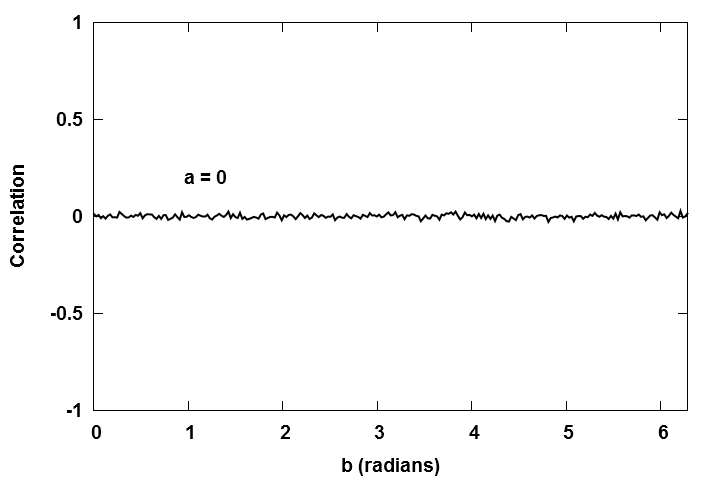}\end{center}

\subsection{Null projection}

When all information transfer is excluded by the EPR conditions, projection of any form cannot occur. To calculate this case, we need only suppose that side B measures the original singlet state that it receives from the source. Following the previous calculations, the steps are obvious and are presented without further comment.

\bigskip
\begin{equation}
\langle A \rangle = Tr([M_A^{2x2}{\otimes}\textbf{1}^{2x2}][\rho_{singlet}^{4x4}]) = 0
\end{equation}

\begin{equation}
P_{A = -1} = P_{A = 1} = 0.5
\end{equation}

\begin{equation}
Sample(P_{i}) \implies O_A \in 1, -1
\end{equation}

\begin{equation}
\langle B \rangle = Tr([\textbf{1}^{2x2}{\otimes}M_B^{2x2}][\rho_{singlet}^{4x4}]) = 0
\end{equation}

\begin{equation}
P_{B = -1} = P_{B = 1} = 0.5
\end{equation}

\begin{equation}
Sample(P_{j}) \implies O_B \in 1, -1
\end{equation}

\begin{equation}
{\langle AB \rangle}_{exp} = \frac{1}{N} \sum_{n=1}^{N} {O_A}_n{O_B}_n \approx 0
\end{equation}

Again, the calculation predicts that no correlation will be observed. The source code of a simulation demonstrating it is available on-line \cite{Graft03}. The form of projection to be used is selectable, that is, between L{\"u}ders projection, Von Neumann projection, and no projection. The result is equivalent to that shown in Figure 2, with only stochastic variation.

\section{Discussion}
\label{Discussion}

L{\"u}ders' rule was developed for the treatment of ensembles \cite{Luders}, so its application to individual projection events is already problematic. Furthermore, by blindly applying L{\"u}ders' rule to physical scenarios for which it is not validly applicable, such as EPR, nonlocality is in effect simply postulated by fiat, whereas L{\"u}ders' rule is in reality not only incorrect for EPR, but is not needed to account for experiments correctly designed and analyzed. Prediction using L{\"u}ders' rule is not a unique necessary quantum mechanical calculation for EPR. Alternative quantum mechanical calculations giving different results are available and required.

A short recounting of some previous work on the topic of this paper is valuable to acknowledge important prior work and to locate my arguments in their proper historical context.

In one of the first reactions to the work of Einstein, Podolsky, and Rosen \cite{Einstein00}, Henry Margenau in 1935 famously rejected the projection postulate in any form for all physical scenarios \cite{Margenau}. He wrote: ``...by the removal of a single postulate commonly accepted, the real difficulty in Einstein-Podolsky-Rosen's conclusion disappears.'' Margenau argued  that no significant quantum calculation requires the projection postulate. Philip Pearle in 1967 also weighed in against the projection postulate \cite{Pearle}, proposing an alternate interpretation of quantum mechanics lacking projection. Pearle did not, however, address the treatment of the EPR paradox to be expected from his alternate interpretation of quantum mechanics. 

Both Margenau and Pearle perhaps go too far in prohibiting projection unconditionally, because we can conceive of physical scenarios where projection may occur. As I have shown here, the correct procedure is to employ the projection rule (including the option of no projection) applicable to the physics of the experiment. For example, a true joint sampling could use L{\"u}ders' rule, while a paradigmatic EPR experiment could not. Importantly, correct projection rules for a given arrangement can be determined empirically.

A recent author that has addressed our topic extensively is Khrennikov \cite{Khrennikov_0,Khrennikov_1,Khrennikov_2,Khrennikov_3,Khrennikov_4}. In \cite{Khrennikov_0} and \cite{Khrennikov_1} Khrennikov argues that Von Neumann projection is crucially different from L{\"u}ders projection and that Von Neumann projection must be used for EPR, dissolving any paradox. Naturally, I am sympathetic to this view, with the caveat that I believe null projection to be more appropriate for EPR. Khrennikov, however, does not address the conflict of this view with the accepted interpretation of the experimental results, whereas I have argued that the experiments, when correctly designed and analyzed, do not show nonlocal correlations.

In \cite{Khrennikov_2} Khrennikov again draws the distinction between Von Neumann and L{\"u}ders projection and claims that neither EPR nor quantum information ``work properly'' when only Von Neumann projection is used. Seemingly contradicting himself, Khrennikov then argues that, in fact, it is a theorem that Von Neumann projection implies L{\"u}ders projection, so there is no problem for the research program of quantum information. Ambiguity continues in \cite{Khrennikov_3}, and as if to correct the contradiction, he cautions that ``conditions of this theorem are the subject of further analysis.'' In \cite{Khrennikov_4} Khrennikov accepts that Von Neumann projection does not entail L{\"u}ders projection.

Khrennikov has done  important work in drawing attention to the role of quantum state projection and the distinction between Von Neumann and L{\"u}ders projection, and for his argument that L{\"u}ders projection is not correct for EPR. Surely, L{\"u}ders projection is not correct for EPR, however, whereas Khrennikov argues that if Von Neumann projection is applied, ``no trace of quantum nonlocality would be found'', I argue here that even Von Neumann projection requires a nonlocal transmission of one bit of information (to signal projection of the singlet state to a mixture), and that null projection is the fully correct way to analyze EPR. I also go much further in asserting  that nonlocal correlations cannot be obtained in EPR experiments, and that the experiments purporting to show nonlocal correlations are incorrectly designed, analyzed, and/or interpreted.

The correct application to EPR of either null projection or Von Neumann projection rather than L{\"u}ders projection blocks quantum correlations and nonlocality, thereby dissolving the paradox. Quantum nonlocality is seen as a mistake. The misapplication of L{\"u}ders projection is seen as the source of apparent nonlocality. We must be careful to apply L{\"u}ders' rule only to appropriate physical scenarios.

The dissolution of the EPR paradox developed here, while thankfully dissolving this terrible paradox that has troubled us for so long, also restores the consistency of our axiom set for physics \cite{Graft00}, allowing for consistent and coherent discourse. 

\renewcommand\refname{References and Notes}


\begin{thebibliography}{99}

\bibitem{Graft00} D. A. Graft, ``On reconciling quantum mechanics and local realism", Proceedings of SPIE conference ``The Nature of Light: What are Photons? 5", SPIE, Bellingham (2013). Also arXiv: quant-ph 1309.1153 (2013).

\bibitem{Graft01} D. A. Graft, ``Analysis of the Christensen et al. Test of Local Realism'', {\it J. Adv. Phys.} {\bf 4}(3), 284-300 (2015).

\bibitem{Einstein00} A. Einstein, B. Podolsky, and N. Rosen, ``Can Quantum-Mechanical Description of Physical Reality Be Considered Complete?" {\it Physical Review},
 {\bf 47}, 777-780 (1935).

\bibitem{Bell00} J. S. Bell, {\it Speakable and Unspeakable in Quantum Mechanics},
2nd ed., Cambridge University Press, Cambridge (1987).

\bibitem{Graft02} \begin{flushleft}
D. A. Graft, the simulation source code is available here: \small \fontfamily{ccr}\selectfont http://rationalqm.us/papers/Rational Interpretation/joint.cpp\end{flushleft} \normalfont \normalsize

\bibitem{Luders} G. L{\"u}ders, ``{\"U}ber die Zustands{\"a}nderung durch den Me{\ss}proze{\ss}'', {\it Ann. Phys. (Leipzig)} {\bf 8}, 322-328 (1951). English Translation by K. A. Kirkpatrick, ``Concerning the state change due to the measurement process'', \textit{Ann. Phys. (Leipzig)}, \textbf{15}, (2006). Also arxiv: quant-ph 0403007 .

\bibitem{Graft03} \begin{flushleft}
D. A. Graft, the simulation source code is available here: \small \fontfamily{ccr}\selectfont http://rationalqm.us/papers/Rational Interpretation/separated.cpp\end{flushleft} \normalfont \normalsize

\bibitem{Pawlowski} Pawlowski, M., Kofler, J., Paterek, T., Seevinck, M., and Brukner, C., ``Non-local setting and outcome information for violation of Bell's inequality'', \textit{New Journal of Physics} \textbf{12}, 083051 (2010). 

\bibitem{no-signaling} Some have argued that information that cannot be extracted is not \textit{really} information, and that superluminal transfer of such ``information'' therefore does not violate special relativity. However, this view is easily refuted. Consider two separated stations A and B. Station A possesses a real variable $a$, and a second randomly selected real variable $r$. Station A generates $b = a + r$ and sends $b$ to station B. Station B cannot extract $a$ from $b$, however, $b$ nevertheless contains information about $a$. This is simply proven because station B can pass $b$ to a third station C, which also receives $r$ from station A. Station C can access $a$ by subtracting $r$ from $b$. It is clear that information about $a$ existed at station B and that the information was passed through station B to station C, despite station B's inability to extract that information. A more intuitive everyday case showing the irrelevance of inability to access the information runs as follows. I write a message and lock it in a box. I send the key to my friend Charlie. I give the box to my friend Bob. Bob possesses the information of the message but cannot access it. We know the information is there from common sense but also because Bob can give the box to Charlie who can open the box and access the information.

\bibitem{Adler} S. L. Adler, D. C. Brody, T. A. Brun, and L. P. Hughston, ``Martingale Models for Quantum State Reduction'', \textit{J. Phys. A} \textbf{34}, 8795-8820 (2001). 

\bibitem{Hegerfeldt} G. C. Hegerfeldt, and R. Sala Mayato, ``Discriminating between the von Neumann and L{\"u}ders reduction rule'', \textit{Phys. Rev. A} \textbf{85}, 032116 (2012). 

\bibitem{VNnote}  H. Margenau, ``Measurements and Quantum States: Part II'',  \textit{Philosophy of Science} \textbf{30}(2) 138-157 (1963). Also, J. Bub, ``Von Neumann's Projection Postulate as a Probability Conditionalization Rule in Quantum Mechanics'', \textit{Journal of Philosophical Logic} \textbf{6}(1), 381-390 (1977). Formally, in EPR the first measurement does not lift the degeneracy and the second measurement can be viewed as the refinement measurement that lifts the degeneracy. Prior to the second measurement, if projection is to occur at all, and if L{\"u}ders projection is excluded, we must view the state at station B prior to the second measurement as a mixture of the possible eigenvalues for the second measurement. Bub has well described the distinction between Von Neumann and L{\"u}ders projection and shown that Von Neumann projection results in a mixture for degenerate eigenvalues.

\bibitem{Margenau} H. Margenau, ``Quantum-Mechanical Description'', \textit{Physical Review} \textbf{49}, 240-242 (1935).

\bibitem{Pearle} P. Pearle, ``Alternative to the Orthodox Interpretation of Quantum Theory'', \textit{American Journal of Physics}, Volume \textbf{35}, Issue 8, 742-753 (1967).

\bibitem{Khrennikov_0} A. Khrennikov, ``EINSTEIN-PODOLSKY-ROSEN PARADOX, BELL'S INEQUALITY, AND THE PROJECTION POSTULATE'', \textit{Journal of Russian Laser Research} \textbf{29}(2), 101-113 (2008).

\bibitem{Khrennikov_1} A. Khrennikov, ``On the problem of completeness of QM: von Neumann against Einstein, Podolsky, and Rosen'', arXiv: quant-phy 0804.2006 (2008).

\bibitem{Khrennikov_2} A. Khrennikov, ``Von Neumann and L{\"u}ders postulates and quantum information theory'', \textit{Int. J. Quantum Information} \textbf{7}, 1303-1311 (2009).

\bibitem{Khrennikov_3} A. Khrennikov, ``Two Versions of the Projection Postulate: From EPR Argument to One-Way Quantum Computing and Teleportation'', \textit{Advances in Mathematical Physics} \textbf{2010}, 945460 (2009).

\bibitem{Khrennikov_4} A. Khrennikov, ``The role of von Neumann and L{\"u}ders postulates in the EPR-Bohm-Bell considerations: Did EPR make a mistake?'', \textit{Int. J. Quantum Information} \textbf{7}(1), 71-81 (2009).

\end{thebibliography}
\end{document}